\documentclass[aps, prb, reprint, superscriptaddress, footinbib]{revtex4-1} 
\usepackage{amssymb}  
\usepackage{amsfonts}  
\usepackage{amsmath}  
\usepackage{amsthm}  
\usepackage{bm}  
\usepackage{bbm}  
\usepackage{xcolor}
\usepackage[colorlinks, allcolors=purple]{hyperref}
\usepackage{graphicx}
\usepackage{tabularx}
\usepackage{multirow}
\usepackage{bbding}
\usepackage{romannum}
\usepackage{comment}

\begin{document}
\title{Quantum entanglement in the $t$-$J$ chain: From charge-spin separation to recombination}

\author{Wayne Zheng}
\affiliation{Institute for Advanced Study, Tsinghua University, Beijing, 100084, China}
\affiliation{Department of Physics, The Ohio State University, Columbus, Ohio 43210, USA}

\author{Zheng-Yu Weng}
\affiliation{Institute for Advanced Study, Tsinghua University, Beijing, 100084, China}

\date{\today}

\begin{abstract}

In contrast to the conventional von Neumann \emph{bipartite} entanglement entropy (bEE), we show that a more appropriate description of the one-dimensional doped Mott insulator is a new kind of \emph{mutual} entanglement entropy (mEE) between the charge and spin degrees of freedom.  
Such a charge-spin mEE 
can clearly distinguish the important and distinct features between the $t$-$J$ model and the so-called $\sigma\cdot{t}$-$J$ model.
In the latter, the phase string sign structure is switched off such that a single doped hole always behaves like a Bloch wave in the whole regime of $J/t$, whereas in the former it exhibits a series of level crossing with the total momentum jumps in the single-hole ground state from spin-charge separation at $J/t\rightarrow 0$ to spin-charge recombination at large $J/t$, which are failed to be detected by bEE. We further show that the distinctions between the two models persist to finite energy density, which can be similarly well characterized by mEE but not by bEE.  
By studying the dynamic time evolution of the states set out of equilibrium at the beginning, we show that mEE indeed always increases with the time, satisfying the common characteristic of entropy.

\end{abstract}

\maketitle
\section{Introduction} 
Quantum entanglement is the correlation of quantum version~\cite{Kitaev20032, RevModPhys.81.865, LAFLORENCIE20161} and stands for the most intrinsic property of quantum systems.
The study of conventional bipartite entanglement entropy (bEE) and its corresponding spectrum has achieved quite a lot in quantum matter physics during recent years, especially for characterizing topological orders~\cite{PhysRevLett.96.110404, PhysRevLett.101.010504, PhysRevB.82.155138, 1508.02595, PhysRevLett.96.110405} and laying a quantum foundation for statistical mechanics~\cite{RevModPhys.91.021001, Nandkishore2015, Luca2016, Popescu2006, Kaufman794}.
A natural question is whether the quantum entanglement can be effectively apply to describing the physics of a quantum many-body system of strongly correlated electrons.

Such an issue has been recently addressed ~\cite{Zheng2018} in the study of the one-dimensional (1D) $t$-$J$ model.
For physical interests, the two-dimensional (2D) $t$-$J$ model is considered as a ``standard model'' closely related to the so-called doped Mott insulator and high-temperature superconductivity in the cuprate~\cite{anderson1997, Baskaran1987, RevModPhys.78.17, Keimer2015}.
Nevertheless, how a doped hole interacts with the surrounding spins even in the 1D $t$-$J$ chain is one of the simplest problem of strong correlation, which still manifests some general physics of the doped Mott insulator. An important discovery in Ref.~\onlinecite{Zheng2018} is that while the bEE fails to capture the rich phase diagram of the one-hole ground state as a function of $J/t$, a new kind of entanglement entropy known as the charge-spin \emph{mutual entanglement entropy} (mEE) can be introduced to characterize the complex phase diagram quite effectively.

\begin{figure}[!h]
    \centering
    \includegraphics[width=.4\textwidth]{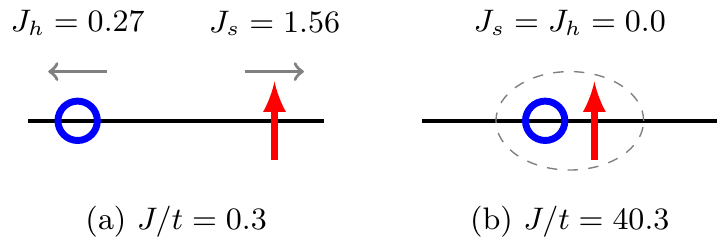}
    \caption{(a) The spin-charge separation is accompanied by nonzero spin and charge currents, $J_{s, h}$, in the 1D $t$-$J$ ring of size $L=10$ at small $J/t$; (b) The charge-spin recombination at larger $J/t$.}
    \label{fig:current}
\end{figure}

In this paper, we shall further examine the physics of the quantum entanglement in the doped Mott insulator by using the simple one-hole-doped spin chain as a toy model.
Here we use both bEE and mEE to comparatively study the $t$-$J$ chain and the so-called $\sigma\cdot{t}$-$J$ chain, respectively.
We find that the $t$-$J$ chain will exhibit rich distinctive features as a function of $J/t$, not only in the ground state but also at finite energy density, which can be all well characterized by mEE. For example, as illustrated in Fig.~\ref{fig:current}(a), at $J/t=0.3$ we find a clear signature of spin-charge separation with separately conserved charge and neutral spin currents at opposite directions in the ground state of a 1D $t$-$J$ ring.
Two degenerate ground states correspond to the reversal of the current flows.
By contrast, at $J/t=40.3$, the charge (hole) and spin are recombined into a quasiparticle as shown in Fig.~\ref{fig:current}(b) without the ground state degeneracy.
On the other hand, in the $\sigma\cdot{t}$-$J$ chain, the charge and spin are always found to be recombined similar to Fig.~\ref{fig:current}(b), in the whole regime of $J/t$.
Such a drastically different ground and excitation states can be also described by mEE.
However, the conventional bEE cannot capture all of these distinctions.  

This paper is organized as follows.
In Sec.~\ref{sec:qe_tjmodel}, we introduce our models and the phase string effect within them.
We compare the mutual entanglement and the conventional bipartite entanglement in these models, proving that the mutual entanglement indeed can reveal more physical results not only in the ground states but also in the excited states.
In Sec.~\ref{sec:time_evolution}, we further study the time-evolution dynamics of these two kinds of quantum entanglement.
Finally in Sec.~\ref{sec:con_dis}, we end up with a brief summary and discussion.

\section{Quantum Entanglement in the $t$-$J$ model}
\label{sec:qe_tjmodel}
\subsection{Phase string effect and the models}
As is well known, the Mottness could dramatically change the Fermi-Dirac statistics, making the conventional Landau-Fermi liquid theory failed in some specific strongly correlated fermionic systems~\cite{Keimer2015, zaanen1599, Weng_2011}.
In particular, one dimensional (1D) interacting fermions generally are described by the Luttinger liquid theory~\cite{tomonaga1950, luttinger1963, Haldane.ll}.
For spin-$1/2$ fermions in the half-filled Mott antiferromagnet limit, once it is doped, it was proposed by Anderson that it will induce a generic \emph{non-local and unrenormalizable quantum phase shift} in the entire Hilbert space~\cite{PhysRevLett.64.1839, PhysRevLett.65.2306} rather than like a quasiparticle's coherent propagation in an effective potential provided by the rest part of the system.
This kind of quantum phase shift has been systematically developed by the so-called phase string theory~\cite{PhysRevB.55.3894, PhysRevB.77.155102}.
One of our goals in this paper is to illustrate such a seemingly abstract mechanism in a much more lucid way with the assistance of quantum entanglement.
Here we take consideration of the simplest but highly non-trivial model for doped Mott antiferromagnets, namely 1D $t$-$J$ model injected with a single hole within the subspace $S=1/2, S^{z}=+1/2$.
The 1D $t$-$J$ Hamiltonian reads $H = H_{t}+H_{J}$ where
\begin{equation}
    \begin{split}
        H_{t} &= -t\sum_{\langle{ij}\rangle, \sigma}(c_{i\sigma}^{\dagger}c_{j\sigma}+h.c.), \\
        H_{J} &= J\sum_{\langle{ij}\rangle}\left(\mathbf{S}_{i}\cdot\mathbf{S}_{j}-\frac{1}{4}n_{i}n_{j}\right).
    \end{split}
    \label{eq:tj}
\end{equation}
It is important to emphasize that the above $t$-$J$ model is meaningful only in the Hilbert space with projecting out the double-occupancy.

To understand the deep consequences of this phase string sign structure, one may introduce a modified the $t$-$J$ model known as the $\sigma\cdot{t}$-$J$ model in which the phase string is precisely ``switched off''~\cite{Zhu201651} for a comparative study. It differs from the $t$-$J$ model only by the hopping term which is given by
\begin{equation}
    H_{\sigma\cdot{t}} = -t\sum_{\langle{ij}\rangle, \sigma}\sigma(c_{i\sigma}^{\dagger}c_{j\sigma}+h.c.)
    \label{eq:sigma_tj}
\end{equation}
while the superexchange term is still $H_{J}$. We shall set the hopping integral $t=1.0$ and vary the ratio $J/t$ with noting that $J/t\simeq{0.3}$ is commonly regarded as close to the realistic situation in the cuprate~\cite{RevModPhys.78.17}.

It has also been previously found~\cite{Zheng2018} that in a finite-size one-hole-doped $t$-$J$ ring, there are a series of distinct ground states characterized by different momenta as one tunes the ratio $J/t$, in contrast to a single ground states of the $\sigma\cdot{t}$-$J$ model in the whole regime of $J/t$.
Figure~\ref{fig:current} shows that for the $t$-$J$ case, the doped hole is charge-spin separated at small $J/t$ but recombined at larger $J/t$, corresponding to these distinct ground states in two limits. In particular, in the spin-charge separation regime, the holon gains a finite current $J_h$, accompanied by a neutral spin backflow current $J_s$.
In other words, the nontrivial total momenta of the ground states found in Ref.~\onlinecite{Zheng2018} can be understood as generated by the nontrivial spin and charge currents, which in turn are due to the phase string effect~\cite{PhysRevB.98.165102} while absent in the $\sigma\cdot{t}$-$J$ model. Here the definition of the currents are given in Ref.~\onlinecite{PhysRevB.98.165102}. Note that different total momenta will correspond to distinct spin and charge currents flowing in opposite directions as a function of $J/t$ (not shown in Fig.~\ref{fig:current}). 

Therefore, the mutual entanglement between the charge and spin degrees of freedom should be crucial to characterize the doped physics in the $t$-$J$ model in contrast to the $\sigma\cdot{t}$-$J$ model.
In the following we  discuss a new kind of mutual entanglement scheme to describe such a doped Mott physics.

\subsection{Entanglement entropy in the eigenstate spectrum}
\begin{figure}[]
    \centering
    \includegraphics[width=.48\textwidth]{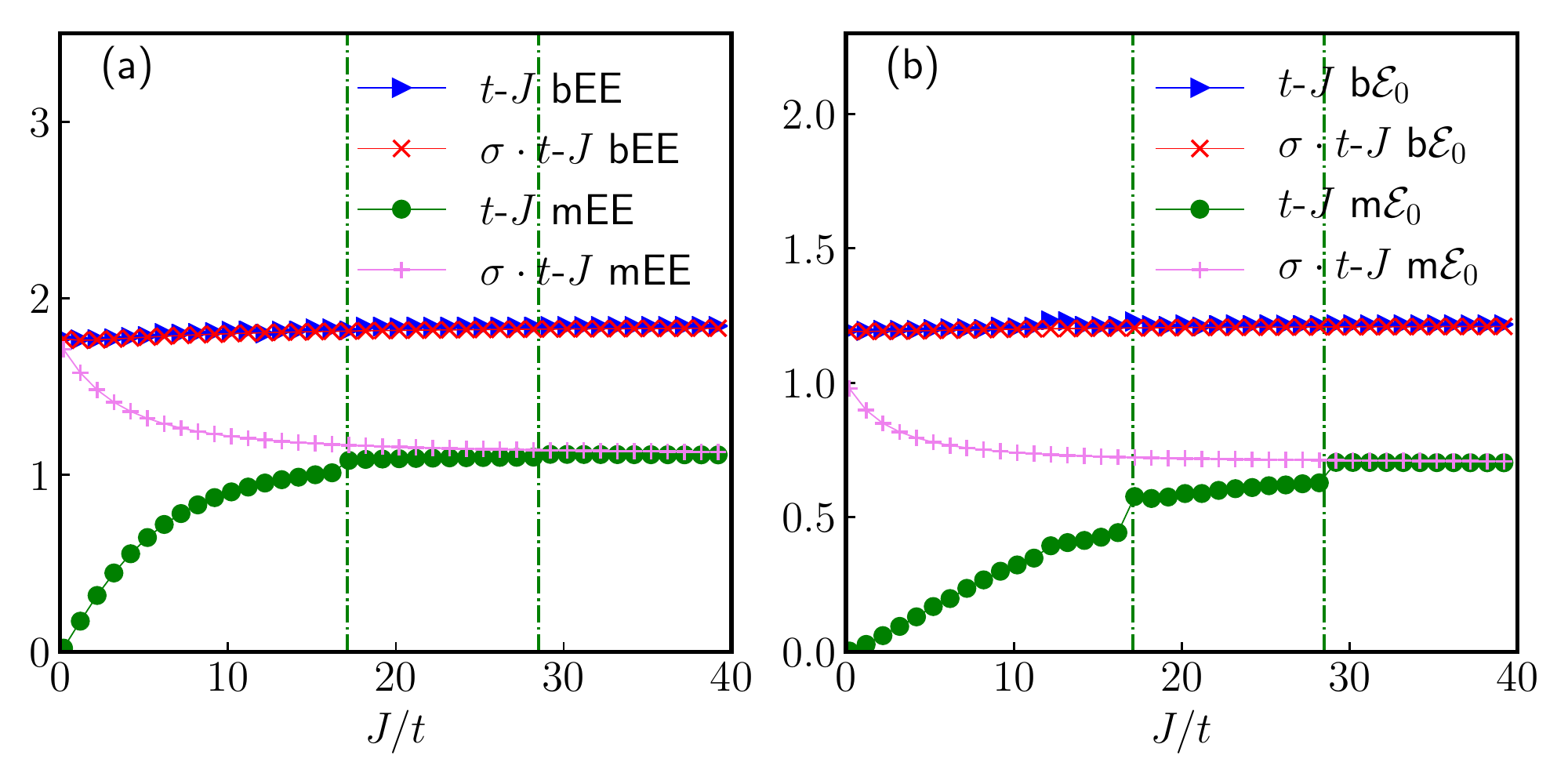}
    \caption{(a) bEE vs. mEE for the single-hole ground states of the $t$-$J$ chain and $\sigma\cdot{t}$-$J$ chain, respectively. While bEE does not distinguish two models, mEE clearly indicates the critical points of $J/t$ at which the ground state of the $t$-$J$ chain shows a momentum jump  (marked by the vertical lines), in contrast to a smooth mEE for the $\sigma\cdot{t}$-$J$ chain which exhibits an increase at small $J/t$ rather than vanishing in the $t$-$J$ case due to spin-charge separation; (b) The lowest eigenvalues $\mathcal{E}_{0}$'s of the corresponding entanglement Hamiltonians show similar behaviors as in (a).}
    \label{fig:both_entanglement}
\end{figure}

Basically, the idea~\cite{Zheng2018} of charge-spin mutual entanglement is based on an operator $P(h)$ to map a one-hole $t$-$J$ configuration $|\alpha\rangle\equiv|h; \{s{'}\}\rangle$ into the direct product of hole position $|h\rangle$ and a spin configuration $|\{s\}\rangle$ as $P(h)|h; \{s{'}\}\rangle=|h\rangle\otimes|\{s\}\rangle$. Note that $s'$ indicates the spin configuration in the original Ising basis of length $L$ in $t$-$J$ model's Hilbert space while $s$ indicates another spin configuration of length $L-1$ in which the hole site is ``squeezed''. Then the wave function can be written as
\begin{equation}
    |\psi\rangle=\sum_{\alpha}v_{\alpha}|\alpha\rangle=\sum_{h, \{s\}}w_{hs}|h\rangle\otimes|\{s\}\rangle ,
\label{direct_prod}
\end{equation}
That is, the original wave function vector $V$ is \emph{reshaped} to a matrix $W$ in the new representation.
By partially tracing out the spin configurations, we can obtain a $L\times{L}$ reduced density matrix $\rho_{h}$ for the hole
\begin{equation}
    \rho_{h}=WW^{\dagger}.
\label{}
\end{equation}
Then the corresponding von Neumann entanglement entropy $S=-\text{tr}(\rho_{h}\ln\rho_{h})$, which is called mEE here, and its entanglement spectrum is also straightforward.
Furthermore, we can also take consideration of the entanglement Hamiltonian~\cite{PhysRevB.83.075102} $\mathcal{H}_{h}$ defined as
\begin{equation}
    \rho_{h}=e^{-\mathcal{H}_{h}}.
    \label{<+label+>}
\end{equation}
In Fig.~\ref{fig:both_entanglement}, we computed both bEE and mEE as well as the entanglement Hamiltonian for the $t$-$J$ chain and $\sigma\cdot{t}$-$J$ chain in the single-hole ground states as a function of $J/t$ on a $L=10$ lattice.
Note that all the numerical results in this paper are obtained by the exact diagonalization method utilizing ARPACKPP~\cite{arpackpp}.
We find that the lowest eigenvalues of the corresponding mutual entanglement Hamiltonians essentially can capture the main feature of the corresponding entanglement entropy, as shown in Figs.~\ref{fig:both_entanglement}(a) and (b).
However, bEE and the corresponding lowest eigenvalue fails to distinguish these different phases, which are also presented in Fig.~\ref{fig:both_entanglement}(a) and (b) for comparison.

\begin{figure}[]
    \centering
    \includegraphics[width=.45\textwidth]{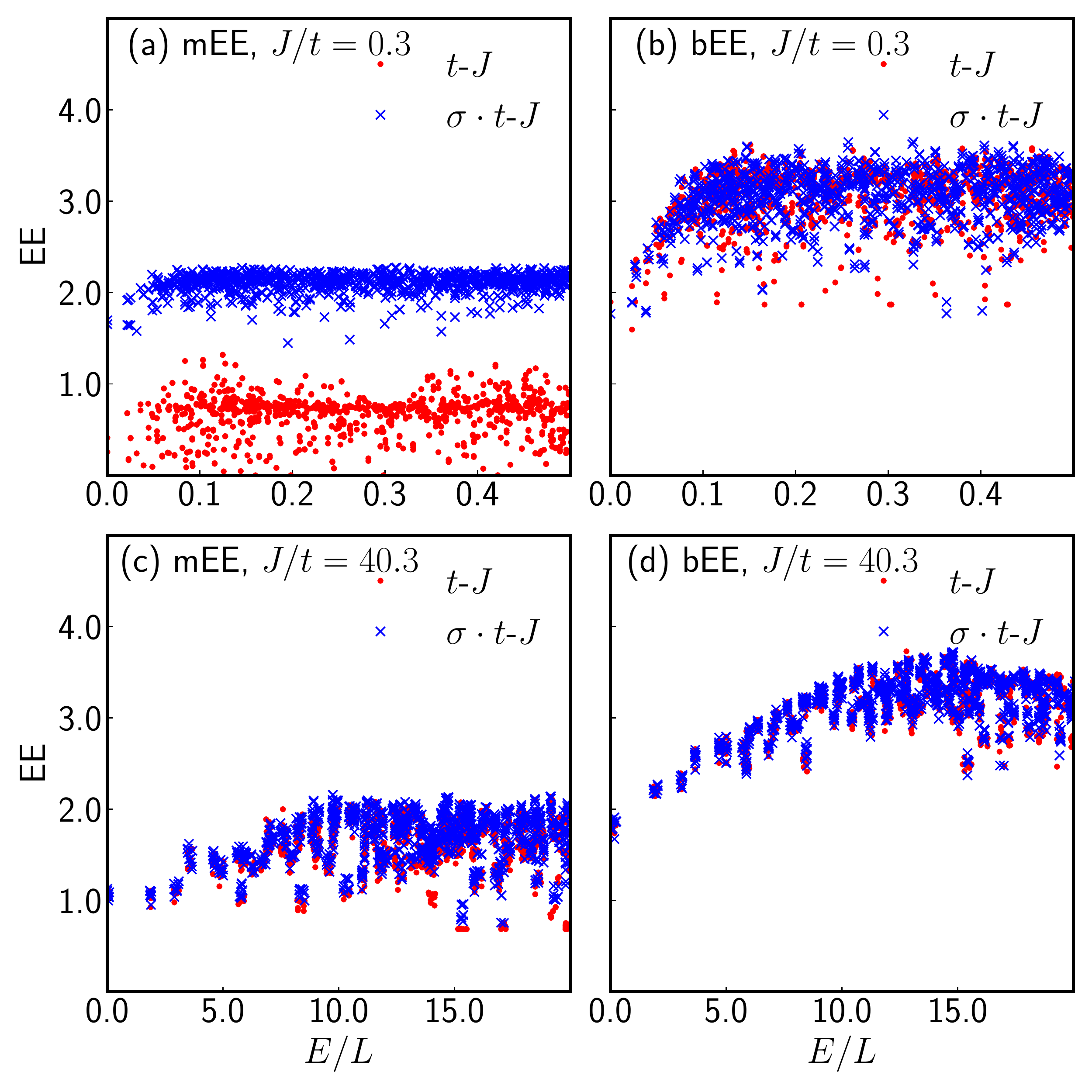}
    \caption{Entanglement entropy for highly excited states of the single-hole-doped $t$-$J$ and $\sigma\cdot$$ {t}$-$J$ chains at $L=10$. The horizontal axis denotes the eigenvalue energy density. Here mEE clearly distinguishes the states of two models at (a) $J/t=0.3$. But two states become indistinguishable at (c) $J/t=40.3$, where they essentially become the same phase in the ground state. By contrast, bEE cannot distinguish any significant distinctions shown by mEE in (b) at $J/t=0.3$ and (c) at $J/t=40.3$.  }
    \label{fig:excited_states_EE}
\end{figure}

In Fig.~\ref{fig:excited_states_EE}, we further compute both bEE and mEE at highly excited eigenstates for both the $t$-$J$ model and $\sigma\cdot$ ${t}$-$J$ model. At $J/t=40.3$, where the spin and charge is recombined as shown in Fig.~\ref{fig:current}(b), both bEE and mEE do not show qualitative difference between the two models as illustrated in Figs.~\ref{fig:excited_states_EE}(c) and~\ref{fig:excited_states_EE}(d). However, at $J/t=0.3$, where the spin-charge separation has been clearly seen in Fig.~\ref{fig:current}(a), we see that mEE clearly indicates the difference between two models not only exhibiting in ground state~\cite{Zheng2018} but also persisting over to the finite-energy density. From Fig.~\ref{fig:excited_states_EE}(b), however, it shows that bEE cannot tell the sharp distinction between the ground states of the two models, one with the spin-charge separation and the other not. 

\begin{figure}[]
    \centering
    \includegraphics[width=.48\textwidth]{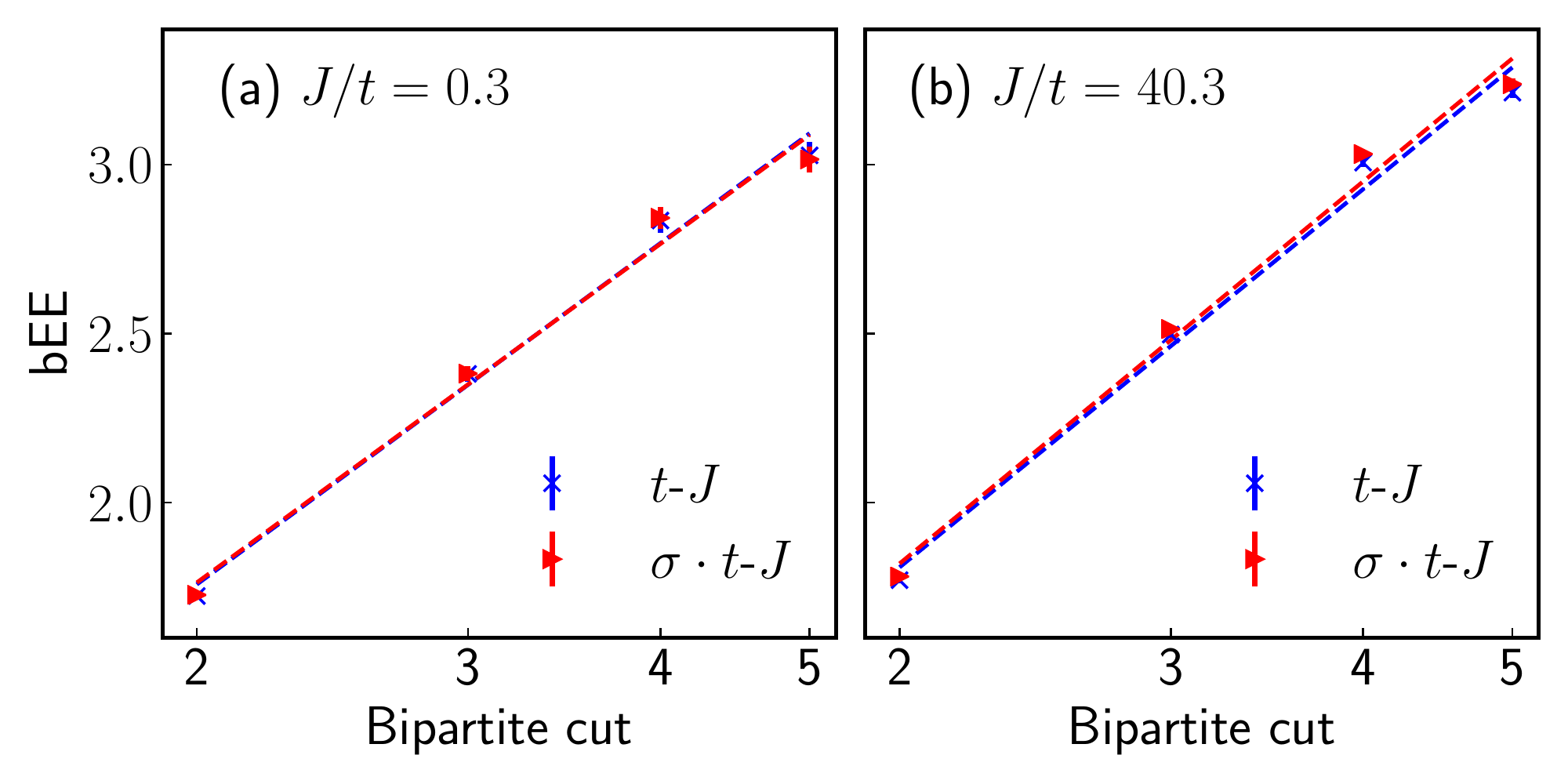}
    \caption{The bEE at different bipartite choices on a $L=10$ lattice, which does not show the distinction between the $t$-$J$ and $\sigma\cdot{t}$-$J$ chains. The horizontal axis is in a logarithmic plot. Here for example, we choose $E$ as the energy of $N=300$ excited state. (a) $J/t=0.3$, $\delta E=0.01$. (b) $J/t=40.3$, $\delta E = 1.0$. ($\Delta E\simeq 5\% E$)}
    \label{fig:bipartite_EE_scale}
\end{figure}

We further examine the scaling behavior of the bEE/mEE in these systems at different bipartite cuts/sample length $L$. First by utilizing the important insight by Wigner~\cite{aprev_chaos} that one should focus on the statistical properties of the spectrum of a many-body system rather than a specific eigenstate, one may introduce the micro-canonical ensemble (MCE) by averaging an operator $O_{\alpha}$ within a relatively narrow energy window $(E-\delta{E}, E+\delta{E})$~\cite{nature06838},
\begin{equation}
    \bar{O}_{mce}=\frac{1}{\mathcal{N}_{E, \delta{E}}}\sum_{\alpha, |E-E_{\alpha}|<\delta{E}}O_{\alpha}
    \label{}
\end{equation}
where $\mathcal{N}_{E, \delta{E}}$ denotes the number of eigenstates within such an energy window. Then we calculate the bEE  based on the MCE rather than in an eigenstate. In Fig.~\ref{fig:bipartite_EE_scale}, the bEE thus calculated roughly obeys a logarithmic behavior, which agrees with the prediction in the fermionic systems~\cite{PhysRevLett.96.010404}. The small deviation is attributed to the finite-size effect and the size of the subsystem approaching half of the total system size. In this sense, the bipartite quantum entanglement in the eigenstates have already widely spread and the so-called eigenstate thermalization hypothesis (ETH)~\cite{PhysRevE.50.888, Nandkishore2015, Luca2016} is valid.
Namely the bEE can be indeed viewed as the thermal entropy of the subsystem. There is no seemingly difference for the all four cases shown in Fig.~\ref{fig:bipartite_EE_scale} no matter with or without the phase string and at small or large $J/t$. 

\begin{figure}[]
    \centering
    \includegraphics[width=.48\textwidth]{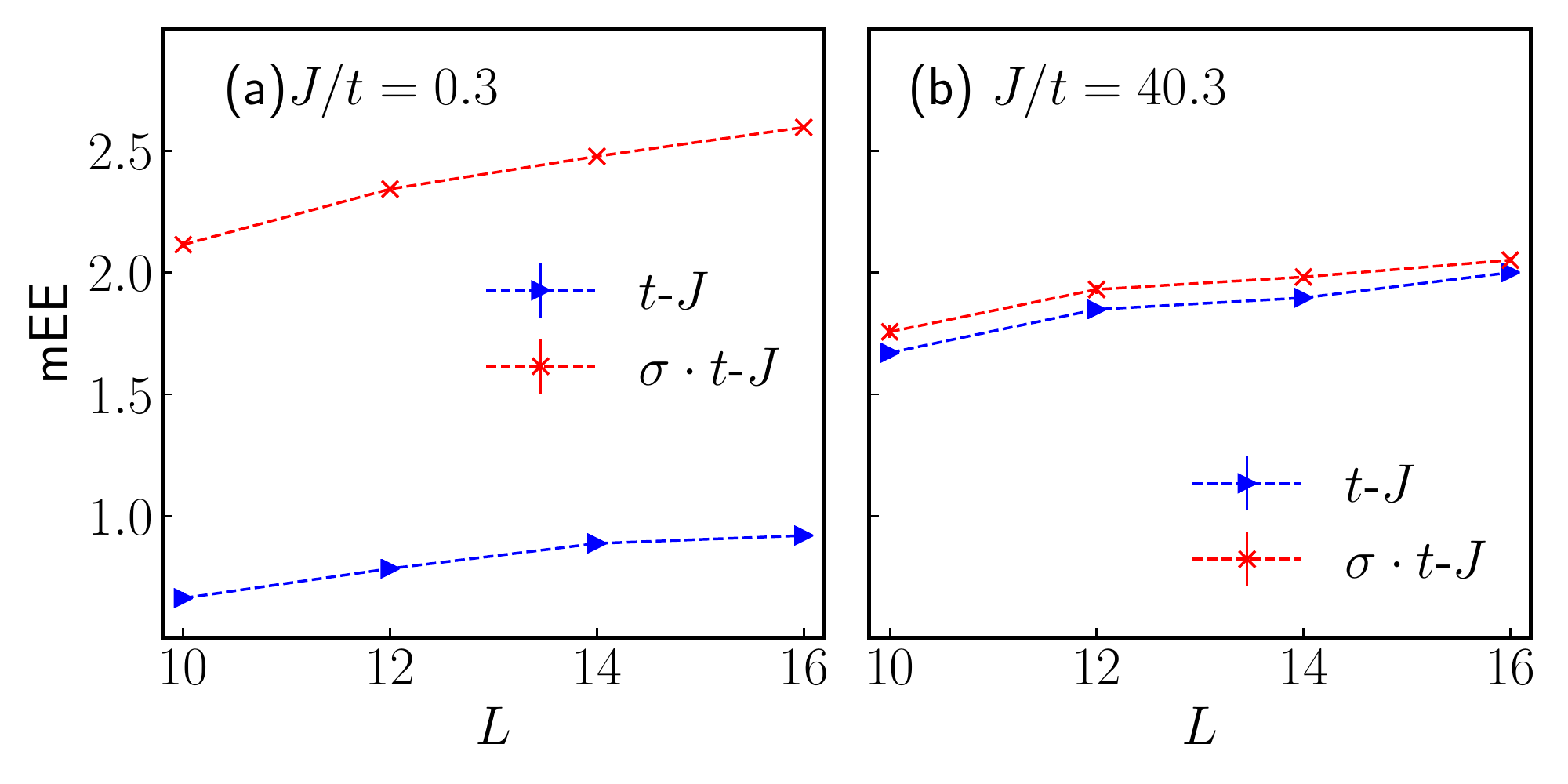}
    \caption{The scaling behavior of the mEE with different lattice sizes $L=10, 12, 14, 16$, which clearly indicates the significant distinction between the $t$-$J$ and $\sigma\cdot{t}$-$J$ chains at $J/t=0.3$, and essentially the same behavior at $J/t=40.3$. Here for example, we choose $E$ as the energy of $N=300$ excited state. (a) $J/t=0.3$, $\delta E=0.01$. (b) $J/t=40.3$, $\delta E = 1.0$. ($\Delta E\simeq 5\% E$)}
    \label{fig:mutual_EE_scale}
\end{figure}

On the other hand, the scaling behavior of the mEE defined on the MCE as illustrated in Fig.~\ref{fig:mutual_EE_scale} clearly indicates the difference between the $t$-$J$ and $\sigma\cdot{t}$-$J$ models at $J/t=0.3$, while displays no obvious distinction at $J/t=40.3$ where the spin-charge recombination is restored. Figure~\ref{fig:mutual_EE_scale}(a) suggests that the charge and spin degrees of freedom are minimally entangled for the $t$-$J$ case with the mEE saturating to a constant at large $L$ because the charge only affects the surrounding spins due to spin-charge separation. But they are maximally entangled in the $\sigma\cdot{t}$-$J$ case at $J/t=0.3$, where the mEE has a similar logarithmic scaling behavior like the bEE as expected. 

\section{Time evolution and the entanglement dynamics}
\label{sec:time_evolution}
\subsection{Quantum chaos scrambling}
\begin{figure}[]
    \centering
    \includegraphics[width=0.45\textwidth]{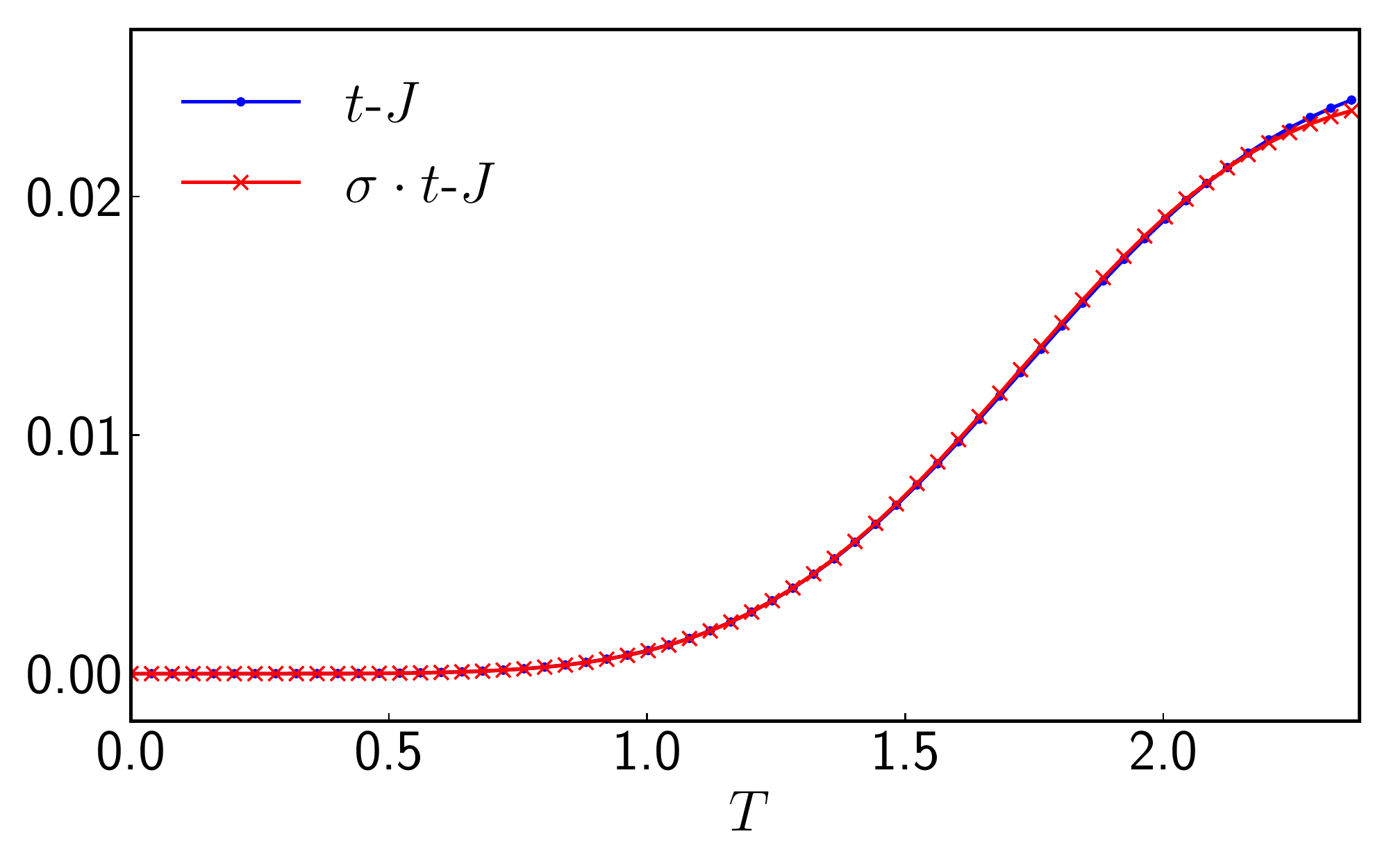}
    \caption{Commutator square $C(T)$ in the early exponential increment period. Here we use $J/t=0.3$. $V$ and $W$ are chosen as the hole density operator $V_{i}=n_{i}^{h}=1-\sum_{\sigma}c_{i\sigma}^{\dagger}c_{i\sigma}$ and $W_{j}=S_{j}^{z}$. It turns out that there is no difference if we choose different $V$ and $W$.}
    \label{fig:butterfly}
\end{figure}

In the first place, here we would like to compute the time dependent square of the operator commutator
\begin{equation}
    C(T)=\langle[V, W(T)]^{2}\rangle_{\beta}\sim{e}^{\lambda_{L}T}
    \label{<+label+>}
\end{equation}
in $t$-$J$ model and $\sigma\cdot{t}$-$J$ model.
It is regarded as the diagnostic of spreading of spatial quantum entanglement and quantum chaos~\cite{Shenker2014, PhysRevLett.115.131603, njp063001}. $\langle\cdot\rangle_{\beta}$ denotes the thermal expectation value and $\lambda_{L}$ is the quantum Lyapunov exponent which reflects how fast chaos develops in a quantum system.
$V$ and $W$ can be chosen as any Hermitian operators which commute at $T=0$.
where $\Lambda$ is the diagonalized Hamiltonian matrix and $P$ is the unitary rotation matrix to diagonalize the Hamiltonian in the original basis.
From Fig.~\ref{fig:butterfly} we can see that the Lyapunov exponents for $t$-$J$ model and $\sigma\cdot{t}$-$J$ model are almost identical.
That is, there is no difference between these two models in terms of the chaos scrambling, which actually is in consistent with our results when it comes to the bEE of these two models since quantum thermalization and chaos scrambling of an isolated quantum system are indeed achieved through the dynamics of bipartite quantum entanglement~\cite{Nandkishore2015, PhysRevLett.88.097905, Kaufman794}.

\subsection{Time evolution of the entanglement entropy}

\begin{figure}[]
    \centering
    \includegraphics[width=0.45\textwidth]{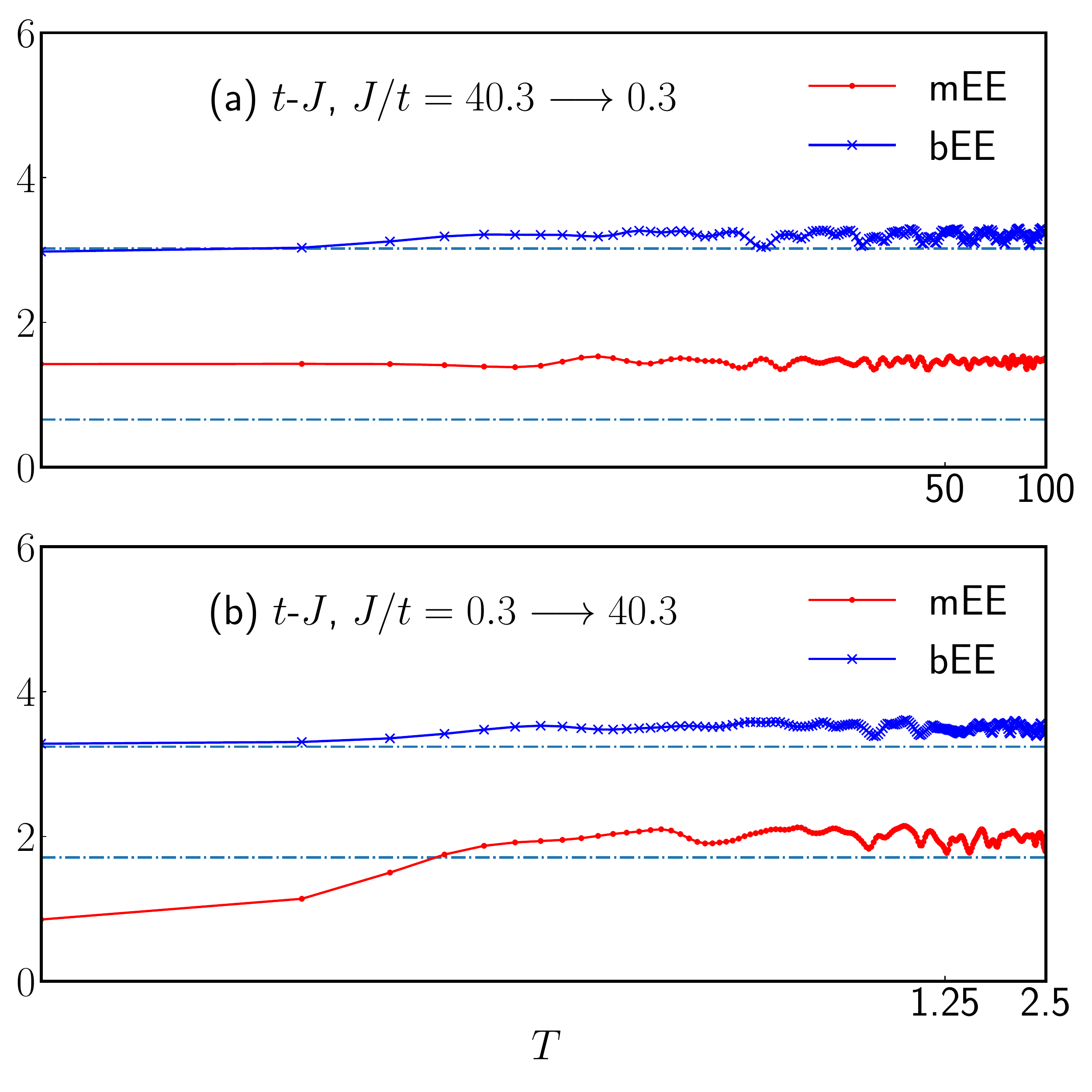}
    \caption{Time evolution of the entanglement entropy (bEE and mEE) of the $t$-$J$ model after a sudden quench of the coupling $J/t$ at time $T=0$: (a) from $J/t=40.3$ to $0.3$; (b) from $J/t=0.3$ to $40.3$. The dash-dotted lines denote the micro-canonical ensemble measurements of the corresponding bEE or mEE. The initial states at $T<0$ are chosen as some arbitrary highly excited eigenstates before quenching.}
    \label{fig:timeEE}
\end{figure}

Furthermore, one may investigate the time evolution (TE) of the entanglement entropy.
We shake the system out of equilibrium by a sudden global quench~\cite{nphys3215} from $J/t=40.3\rightarrow0.3$ or from  $J/t=0.3\rightarrow 40.3$ at time $T=0$. Then we follow the time evolution of the entanglement entropy (bEE and mEE) at $T>0$. The dimensionless number $Tt$ will be taken up to $100.0$ for $J/t=0.3$ and to $2.5$ for $J/t=40.3$, respectively, to reach the saturation of the entropy due to the difference in $J/t$ (cf. Fig.~\ref{fig:timeEE}). 

In Fig.~\ref{fig:timeEE}(a), the dash-dotted lines indicate the bEE and mEE in the equilibrium at $J/t=0.3$ measured by MCE. There is a strong deviation of the TE mEE even after a long-time evolution, but the TE bEE remains rather close to the equilibrium line. Here the corresponding initial states at $T<0$ are chosen from some arbitrary highly excited eigenstates. Table~\ref{tab:quench_small} further shows three examples of different initial states chosen arbitrarily from the eigenstates before quenching, with comparing the saturated TE measurement and MCE measurement of the bEE and mEE for both $t$-$J$ and $\sigma\cdot{t}$-$J$ models.

Thus, for the $t$-$J$ system to evolve from the spin-charge recombined phase at $J/t=40.3$ to the spin-charge separation phase at $J/t=0.3$, the TE of the mEE shows a large deviation from the final equilibrium state but bEE does not. As for the $\sigma\cdot{t}$-$J$ model, both mEE and bEE show convergence of the TE to the MCE. All of these are consistent with the notion of an entropy, for both bEE and mEE, that monotonically increases with the time (after a coarse-grain average) and saturates in an equilibrium state. The large deviation of the TE of the mEE is simply due to the fact that the MCE value at $J/t=0.3$ is much smaller than that at $J/t=40.3$ and the TE of the mEE can never be reduced to the latter in the large $T$.

For comparison, we also present the TE of the mEE and bEE for the $t$-$J$ case from  $J/t=0.3\rightarrow 40.3$ at time $T=0$ in  Fig.~\ref{fig:timeEE}(b). In this case, both bEE and mEE in TE saturate to the values of the MCE in the large $T$. Similarly both TE values in large $T$ have been shown in Table~\ref{tab:quench_large} at three different eigenstates before quenching for both $t$-$J$ and $\sigma\cdot{t}$-$J$ models. Note that for the $\sigma\cdot{t}$-$J$ model, the TE of the mEE has shown a discrepancy from the MCE at large $T$ due to the fact that the MCE value is \emph{smaller} at $J/t=40.3$ than at $0.3$. Again, the non-decreasing property of the TE of the mEE is at working.

\begin{table}
    \caption{The mEE and bEE measured by the TE and MCE after quenching from $J=40.3\rightarrow0.3$ for the $t$-$J$ and $\sigma\cdot{t}$-$J$ models at three different initial excited eigenstates.\\}
    \centering
    \begin{ruledtabular}
        \begin{tabular}{p{.04\textwidth}p{.08\textwidth}p{.08\textwidth}p{.08\textwidth}p{.08\textwidth}p{.08\textwidth}}
            &  & \multicolumn{2}{c}{$t$-$J$} & \multicolumn{2}{c}{$\sigma\cdot{t}$-$J$} \\
        &  & mEE & bEE & mEE & bEE \\
        \hline
        \multirow{2}{.04\textwidth}{(a)} & TE & {1.98(9)} & 3.52(1) & 2.11(1) & 3.53(7) \\
        & MCE & {0.66(3)} & 3.02(7) & 2.11(5) & 3.01(5) \\
        \hline
        \multirow{2}{.04\textwidth}{(b)} & TE & {2.01(5)} & 3.51(1) & 2.10(4) & 3.47(0) \\
        & MCE & {0.67(7)} & 3.07(8) & 2.10(7) & 3.09(0) \\
        \hline
        \multirow{2}{.04\textwidth}{(c)} & TE & {1.94(9)} & 3.77(3) & 2.22(2) & 3.65(2) \\
        & MCE & {0.68(1)} & 3.08(5) & 2.09(6) & 3.04(5) \\
    \end{tabular}
    \end{ruledtabular}
    \label{tab:quench_small}
\end{table}

\begin{table}
    \caption{The mEE and bEE measured by the TE and MCE after quenching from $J=0.3\rightarrow40.3$ for the $t$-$J$ and $\sigma\cdot{t}$-$J$ models at three different initial excited eigenstates.\\}
    \centering
    \begin{ruledtabular}
        \begin{tabular}{p{.04\textwidth}p{.08\textwidth}p{.08\textwidth}p{.08\textwidth}p{.08\textwidth}p{.08\textwidth}}
            &  & \multicolumn{2}{c}{$t$-$J$} & \multicolumn{2}{c}{$\sigma\cdot{t}$-$J$} \\
        &  & mEE & bEE & mEE & bEE \\
        \hline
        \multirow{2}{.04\textwidth}{(a)} & TE & 1.99(9) & 3.57(5) & 2.03(2) & 3.00(6) \\
        & MCE & 1.70(7) & 3.24(0) & 1.74(6) & 3.26(4) \\
        \hline
        \multirow{2}{.04\textwidth}{(b)} & TE & 1.88(5) & 3.54(0) & 2.22(1) & 3.54(5) \\
        & MCE & 1.72(9) & 3.29(1) & 1.77(8) & 3.30(6) \\
        \hline
        \multirow{2}{.04\textwidth}{(c)} & TE & 1.62(9) & 2.88(9) & 2.21(3) & 3.49(6) \\
        & MCE & 1.61(1) & 3.20(4) & 1.69(8) & 3.20(9) \\
    \end{tabular}
    \end{ruledtabular}
    \label{tab:quench_large}
\end{table}

\section{Conclusion and discussion}
\label{sec:con_dis}

In this work, we explored the quantum entanglement description of strong correlation in the one-hole-doped $t$-$J$ chain by using ED.
We examined two kinds of entanglement entropy, namely, the conventional von Neumann bipartite entanglement called bEE and the mutual entanglement between the charge and spin degrees of freedom called mEE introduced in Ref.~\cite{Zheng2018}. 
Our results clearly showed that whereas bEE fails to detect the distinct phases as a function of the ratio $J/t$, including the spin-charge separation as the hallmark of strong correlation in the small $J/t$ regime of the  $t$-$J$ model, mEE can effectively identify all of them, not only in the ground state, but also in highly excited states of finite energy density.
In particular, we made a comparative study of the $t$-$J$ model with the so-called $\sigma\cdot t$-$J$ model, in which the phase string is turned off to result in a more conventional (Bloch wave like with the spin-charge recombination) behavior of the doped hole. As expected, mEE clearly distinguishes the two models but bEE cannot.
Furthermore, the time-evolution of the out-of-equilibrium states behaves differently in different regimes, which can be still well characterized by mEE due to its entropy-like property, whereas bEE is not sensitive at all.

Based on the mEE description presented above, the one-hole-doped $t$-$J$ and $\sigma\cdot{t}$-$J$ models are most distinct at small $J/t$ limit, where mEE vanishes for the first while it reaches the maximum for the second [cf. Fig.~\ref{fig:both_entanglement}(a)]. 
Namely the charge and spin are indeed separated in the $t$-$J$ model but are most strongly entangled in the $\sigma\cdot{t}$-$J$ model at $J/t\ll 1$. In the undoped Heisenberg chain, spins are long-range correlated such that each spin has a maximal amount of mutual entanglement with the other spins, which would remain the same if the spin is replaced by a hole which follows a similar dynamics as the original spin. This should be the case for the $\sigma\cdot{t}$-$J$ model or the $t$-$J$ model at $J/t\gg 1$. We call these charge-spin recombined states.
In the spin-charge separation regime of the first model at the small $J/t$, however, the holon as the dressed hole will carry a momentum (charge current) while generate a neutral backflow spin current (cf. Fig.~\ref{fig:current}) due to the phase string effect, whose magnitudes are dependent on $J/t$ and such a mutual influence between the two degrees of freedom is well captured by mEE.
Therefore, the mEE can provide a precise and effective description of a doped hole strongly correlated with the background spins in a 1D closed loop system.
How to generalize the present approach to the ladder or two-dimensional case, or to finite doping will be highly interesting to explore in the future.

To identify the generic and intrinsic property of a doped Mott insulator, strange metal regime seems to be the optimal candidate rather than, say, pseudogap or superconductivity regimes because there always exists some other competing orders which may mislead our understanding route.
The fact that strange metal is regarded as an ideal realization of quantum matter without quasiparticles has been widely accepted and studied~\cite{1612.07324, PhysRevB.89.245116} for a long time.
From Fig.~\ref{fig:excited_states_EE} we can see that the disentangled feature of the two degrees of freedom in $t$-$J$ model are robust and always persist in highly excited states corresponding to a high-temperature strange metal regime, which can provide a good route towards understanding strange metals.

The idea of mutual entanglement can also be generalized to other situations, for instance, the Hubbard model.
Suppose there are $M$ sites comprised of the lattice and $N_{\sigma} (\sigma=\uparrow, \downarrow)$ electrons ($N_{\uparrow}+N_{\downarrow}\leqslant{2M}$).
The dimension of its Hilbert space is $\mathfrak{d}=\prod_{\sigma}C_{M}^{N_\sigma}$.
It is convenient to use the tensor product of spin-up and -down electrons' Hilbert space to fuse the Hilbert space of Hubbard model.
Therefore the wavefunction can be written as a matrix~\cite{PhysRevLett.62.1201} $|\psi\rangle_=\sum_{\alpha\beta}W_{\alpha\beta}|\alpha\rangle_{\uparrow}\otimes|\beta\rangle_{\downarrow}$.
The mutual entanglement between spin-up and pin-down electrons can be defined naturally and discussed as the route above.
Its physical interpretation is left in the future.

The authors acknowledge very helpful discussions with Y.M. Lu, X.L. Qi and Y.F. Gu.
This work is partially supported by Natural Science Foundation of China (Grant No. 11534007), MOST of China (Grant Nos. 2015CB921000 and 2017YFA0302902).

\appendix

\section{Time evolution approximation}

\begin{figure}[]
    \centering
    \includegraphics[width=0.48\textwidth]{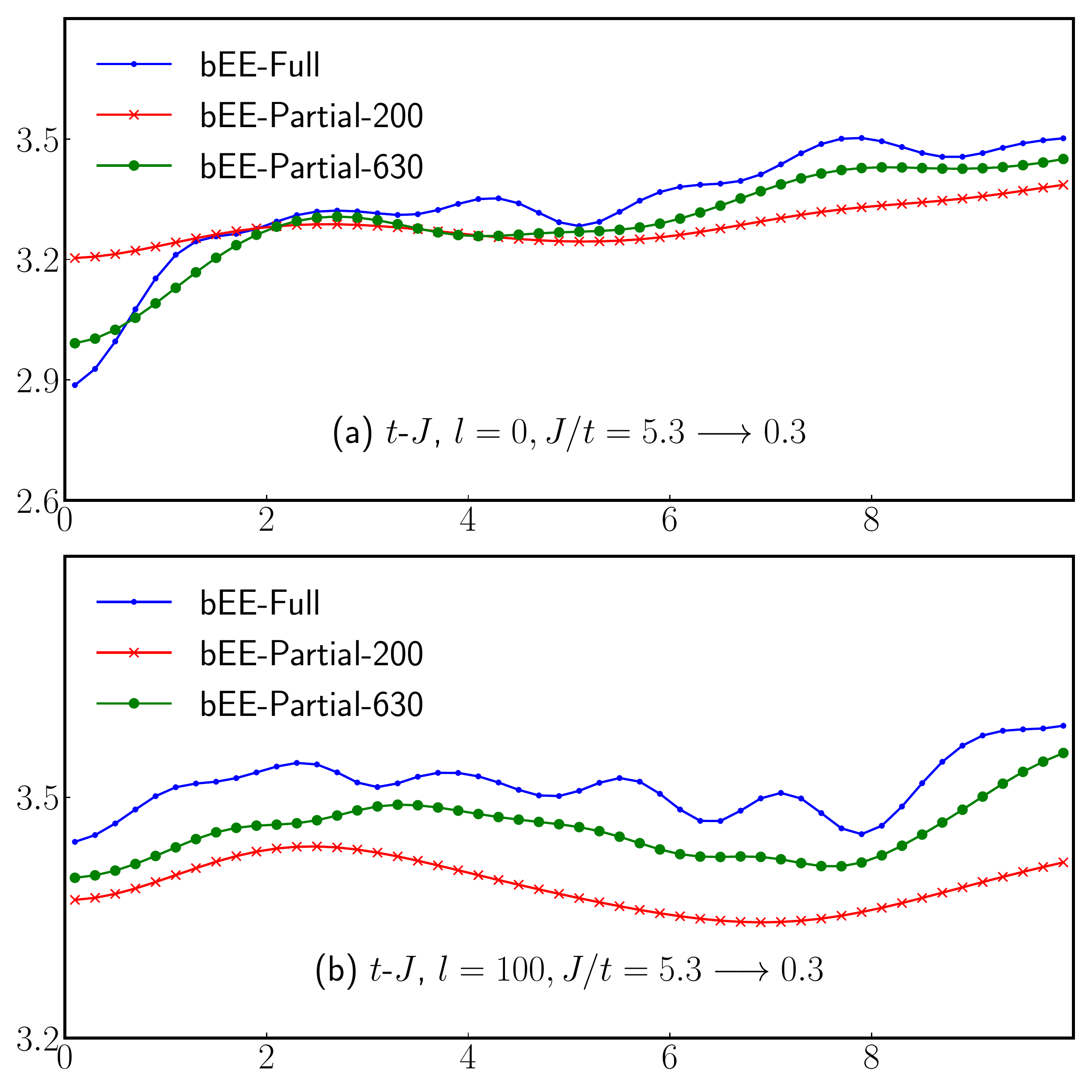}
    \caption{Approximated time evolution test with bEE.}
    \label{fig:test_te}
\end{figure}

Generally speaking, a Hamiltonian matrix in a specific representation can be diagonalized in the form as
\begin{equation}
    P^{\dagger}HP=D,
    \label{}
\end{equation}
where $D$ is the diagonal matrix and $P$ is the unitary transformation consisting of all the eigenvectors of $H$.
In another way, with $HP=PD$ explicitly written as
\begin{equation}
    H(\mathbf{p}_{0}, \cdots, \mathbf{p}_{N-1})=(\lambda_{0}\mathbf{p}_{0}, \dots, \lambda_{N-1}\mathbf{p}_{N-1}),
    \label{<+label+>}
\end{equation}
where $P$ is written as a column vector array and $\lambda$s are the eigenvalues of $H$. That is, $H\mathbf{p}_{i}=\lambda_{i}\mathbf{p}_{i}, i=0, \cdots, N-1$.
$N$ is the dimension of the Hilbert space. Then the time evolution operator can be written as
\begin{equation}
    U(T)=e^{-\text{i}HT}=e^{-\text{i}PDP^{\dagger}T}=Ue^{-\text{i}DT}U^{\dagger}
    \label{<+label+>}
\end{equation}
for the sake of $P$ is unitary.
A practical issue here is that the exact $P$ matrix requires the full spectrum while the ARPACKPP~\cite{arpackpp} package which uses the kind of iteration algorithm is much more time consuming when it comes to requiring higher and higher eigenvalues as well as the corresponding eigenvectors.
A possible kind of method is to keep only the lowest $M (M<N)$ eigenvalues to approach an approximation as good as possible if we just involve with a relatively low-energy ensemble or pure states.
\begin{equation}
    H=PDP^{\dagger}\simeq\sum_{i=0}^{M-1}\lambda_{i}\mathbf{p}_{i}\mathbf{p}_{i}^{\dagger}.
    \label{<+label+>}
\end{equation}
Therefore, time evolution operator can be approximated as
\begin{equation}
    U(T)=e^{-\text{i}HT}\simeq\sum_{i=0}^{M-1}e^{-\text{i}\lambda_{i}T}\mathbf{p}_{i}\mathbf{p}_{i}^{\dagger}.
    \label{}
\end{equation}
Choosing $M$ states may depend on the thermodynamic ensemble temperature $\beta$ and other related factors.
In Fig.~\ref{fig:test_te} we did a simple test in terms of the time evolution of bEE.
It shows that if we kept more and more eigenvectors to construct the time evolution operator, it indeed keeps its dynamical track better and better.
It also shows that higher excited states entering into the time evolution operator can detect higher frequency as well as more sensitive time evolutionary dynamics.

\bibliographystyle{apsrev4-1}
\bibliography{quantumEntanglement_tJchain_ref}
\end{document}